\documentclass[conference]{IEEEtran}
\IEEEoverridecommandlockouts
\usepackage{cite}
\usepackage{amsmath,amssymb,amsfonts}
\usepackage{algorithmic}
\usepackage{graphicx}
\usepackage{textcomp}
\usepackage{xcolor}
\usepackage{array}
\def\BibTeX{{\rm B\kern-.05em{\sc i\kern-.025em b}\kern-.08em
    T\kern-.1667em\lower.7ex\hbox{E}\kern-.125emX}}

\begin{document}

\title{Toward Edge-enabled Cyber-Physical Systems Testbeds}


\author{\IEEEauthorblockN{V. K. Cody Bumgardner}
\IEEEauthorblockA{Department of Computer Science\\
University of Kentucky\\
Lexington, Kentucky, USA\\
Email: cody@uky.edu}
\and
\IEEEauthorblockN{Nima Seyedtalebi}
\IEEEauthorblockA{Department of Computer Science\\
University of Kentucky\\
Lexington, Kentucky, USA\\
Email: nima.seyedtalebi@uky.edu}
\and
\IEEEauthorblockN{Caylin Hickey}
\IEEEauthorblockA{Department of Computer Science\\
University of Kentucky\\
Lexington, Kentucky, USA\\
Email: caylin.hickey@uky.edu }}

\maketitle

\begin{abstract}

The use of edge computing can be extremely valuable in support of CPS efforts.  However, few if any testbeds provide the type of resource control and provisioning required to support edge-enabled CPS experimentation.  Likewise, commercial offerings provide operational capabilities, but lack the distributed infrastructure and transparency provided by research testbed.  In this paper we propose methods to develop new and augment existing testbeds to better support the challenges of edge computing and CPS research.  The proposed network is specifically designed to address the challenges associated with edge-based provisioning, data collection, analysis, monitoring, and measurement across islands of edge and data center resources.    

We present the purpose of our work, the basic architecture, initial results, the relationship to the existing software, and the potential of an existing edge-focused framework to support the foundations of edge-focused CPS testbeds.

\end{abstract}


\IEEEpeerreviewmaketitle

\section{Introduction}

As society comes to depend on the Internet of Things (IoT), the need for stable and secure environments supporting the development and testing of Cyber-Physical Systems (CPS) \cite{baheti2011cyber}, Machine to Machine (M2M) \cite{boswarthick2012m2m} technologies, Industrial Internet \cite{evans2012industrial}, and Smart City \cite{zanella2014internet} applications and related operational platforms becomes critical. 

In recent years a number of software platforms \cite{web:awsiot,web:thingworx,web:xively,web:boschiot,web:sensorcloud,web:iotivity} have been developed to support IoT efforts.  The majority of these platforms rely on remote and potentially distant public cloud providers such as Amazon EC2 \cite{amazon2010amazon} or Microsoft Azure \cite{web:azure} for infrastructure.  A number of vendors provide for commercial support and production quality hosting for device-cloud operations, where remote devices communicate directly with cloud-based infrastructure.  However, there are cases (latency, security, locality, etc.) \cite{cyber2016framework} where it is more appropriate or necessary to move functionally closer to sources of data generation, as common in so-called fog \cite{bonomi2012fog} and edge \cite{shi2016edge} computational paradigms.  The authors believe that the importance of edge computing will only increase with the size and complexity of CPS networks.  The coordination of edge communications, shared resources, high-level application scheduling, monitoring, measurement, and Quality of Service (QoS) enforcement are just a few areas that must be further developed to realize the benefits of edge computing in support of CPS.        

The Global Environment for Networking Innovation (GENI) \cite{berman2014geni}, Future Internet Research and Experimentation (FIRE) \cite{gavras2007future}, and more recently CloudLab \cite{ricci2014introducing}, are the some of most prominent next generation network and innovation testbed projects.  These projects aim to support ``at scale" experimentation using a network of globally distributed resources.  Unlike resources distributed by commercial public cloud providers in massive data centers, the aforementioned testbed projects distribute smaller groups of resources between large numbers of locations.  For example, GENI resources are distributed across the US and beyond on college campuses, research laboratories, and within local city governments.  The GENI-type distribution of infrastructure allows computational resources to be applied to data sources much closer than public cloud providers.      

In efforts to address the specific CPS experimentation needs \cite{sanchez2011smartsantander}, a number of independent IoT testbeds \cite{grace2015analysis} have been deployed in federation with FIRE across Europe.  The majority of these testbeds focus on lower-level communications or device data collection and typically range in scope from building to city-scale experimentation.  In the US, CPS and smarter cities applications, like those developed as part of US Ignite \cite{ricart2014us} efforts, are often deployed making used of low-level GENI, CloudLab, or other cloud-based resources.  

The need for edge computing, especially in relation to CPS and smarter cities efforts, has made innovation environments like GENI attractive options for experimental application development and testing.  However, while these experimental resources are suitable as testbeds, they were never intended to function as an operating platform for high-level applications, especially those that require significant and stable infrastructure with end-to-end resource measurement, monitoring, and Quality of Service (QoS) enforcement.  This is not to say that existing testbeds are not useful for edge and CPS experimentation, just that the experimenters must develop their own methods, compensating controls, and estimations to address services that might be better provided on the testbed platform-level.    

The following items must be addressed in the development of new or augmentation of existing testbeds for the purposes of edge-enabled CPS experimentation: a) Stability of computing and network resources; b) Support and management a large number of objects; c) QoS enforcement of resource reservations; d) End-to-end monitoring and measurement of resources; e) Application-level experimentation specifications and deployment services.  While other requirements may exist to support edge-enabled CPS experimentation, the  topics covered have presented challenges to the authors and to the best of our knowledge are not currently addressed by existing testbeds or commercial software packages.  While many operational requirements might be satisfied by existing commercial offerings such as AWS IoT Greengrass \cite{awsgreengrass}, Azure IoT Edge \cite{microsoftiot}, and Google Cloud IoT \cite{googleiot}, such offerings do not negate the need for testbed environments.

A gap exists between production cloud-based IoT frameworks, which are often focused on consumer devices, existing testbeds, which often focus on low-level infrastructure provisioning, and the type of services required to support edge-enabled experimentation in the area of CPS.  While this paper does not claim to close the described gap, we do propose a number of challenges that could serve to advance the field if addressed in new or existing testbeds. Thus, our contribution is the enumeration of the challenges we and other researchers have faced in practice and a discussion of how they might be addressed in future testbeds.

In full disclosure, the authors were involved in the support of the GENI network, so discussion will be heavily influenced by experience in GENI operations.  In addition, much of the proposed architecture and initial results will be framed in the context of Cresco \cite{7818455,8480196}, an edge computing framework developed by the authors, and lessons learned in the development and support of a number of edge computing projects \cite{Bumgardner:2014:SHS:2568088.2568103, 7502958, 8480233, 7749534, crescosmartcity, crescodefense, crescohomeland, cresconetdemo, crescodistributed, crescoai, crescogenomic}. 

The remainder of this paper will cover a proposed architecture and related experimental results towards the development of edge-enabled CPS testbeds.

\section{Architecture}

While the service-level requirements of production environments differ from experimental testbeds, edge-enabled CPS experiments can be deployed in the existing testbeds, such as the GENI environment.  Using bare metal provisioning and software-defined networking \cite{mckeown2009software} capabilities of the GENI network, we can stitch together islands of low-level edge resources, which are indistinguishable from standalone resources from the infrastructure standpoint.  The remainder of this paper describes platform services that could be developed as part of a new testbed or that could work from within, or adjacent to, GENI resources depending on the desired service-level requirements.  Our focus will be on how we might augment the existing GENI environment making use of low-level infrastructure ``slicing" and resource provisioning.  Unless otherwise noted, we will make use of Cresco to interface existing or new infrastructure with edge-enbled services.  The Cresco framework was created to address the challenges of managing resources across local (campus), regional (state), and global (country/world) domains.  As is common with software-defined systems, including GENI, we separate the control and application planes.  While the Cresco framework is used to provide the control plane, additional frameworks and/or federations managed by Cresco can be used on or provide the data and application planes.  

We propose the following architecture components to support edge-enabled CPS testbeds: 

\subsection{Operational Stability} Testbed environments provide low-level access to physical and virtual infrastructure, which is needed for experimentation of things such as protocol and device development.  In testbeds such as GENI, provisioning of low-level infrastructure must be coordinated between heterogeneous hardware and software implementations, which are geographically distributed, at times around the globe.  The testbed scheduling services often have little or no information pertaining to the operational state of the underlying system.  This type of high-level scheduling of low-level, geographically distributed resources is very different from the way cloud providers, such as Amazon EC2 provide resources.  For example, a cloud provider has complete control over their underlying infrastructure and software stack used in the provisioning of virtual resources.  In comparison, GENI must manage the the low-level stitching of communication paths through Internet2 in conjunction with local (campus) networks, and the provisioning of computational resources across heterogeneous environments.  These scheduling practices, while necessary for experimentation and testbed environments, lead to high rates of provisioning failures, with a stable post-provisioning steady-state.  A successful CPS testbed must focus on providing stable environments for CPS applications over flexibility of the underlying infrastructure.  Within this context, it is preferable to have a stable environment for controlled simulation than the flexibility to provision physical devices.   We are not taking the physical out of CPS, but the physical attributes are so diverse that we can only hope to support experimentation through virtual infrastructure and simulation.  Once core underlying resources have been provisioned, they remain online as long as the framework is active.  On the infrastructure level, the edge-enabled CPS services themselves appear  as an experiment.  Applications should share underlying low-level resources managed and monitored by the requesting edge framework.  The proposed architecture does not prohibit access to low-level hardware such as radios, sensors, et cetera; from an edge computing perspective, access to these devices can be gained through external IoT gateways, lower-level framework federations, or directly over higher-level protocols.       

\begin{figure}[ht!]
  \centering
      \includegraphics[width=.4\textwidth]{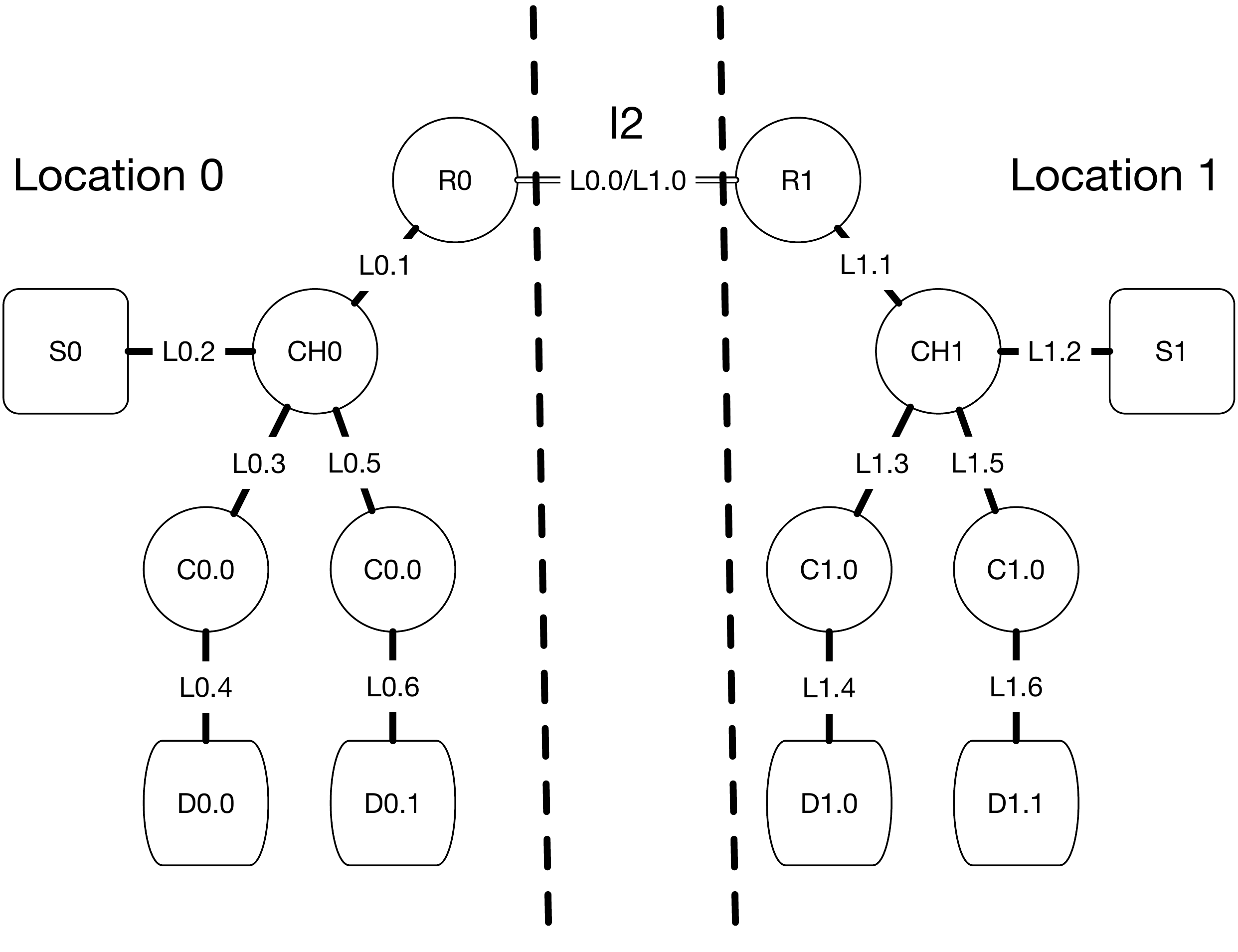}
  \caption{Edge-enabled Resources Connected over Internet2}
  \label{fig:sites_over_i2}
\end{figure}

\begin{figure}[h!]
  \centering
      \includegraphics[width=.4\textwidth]{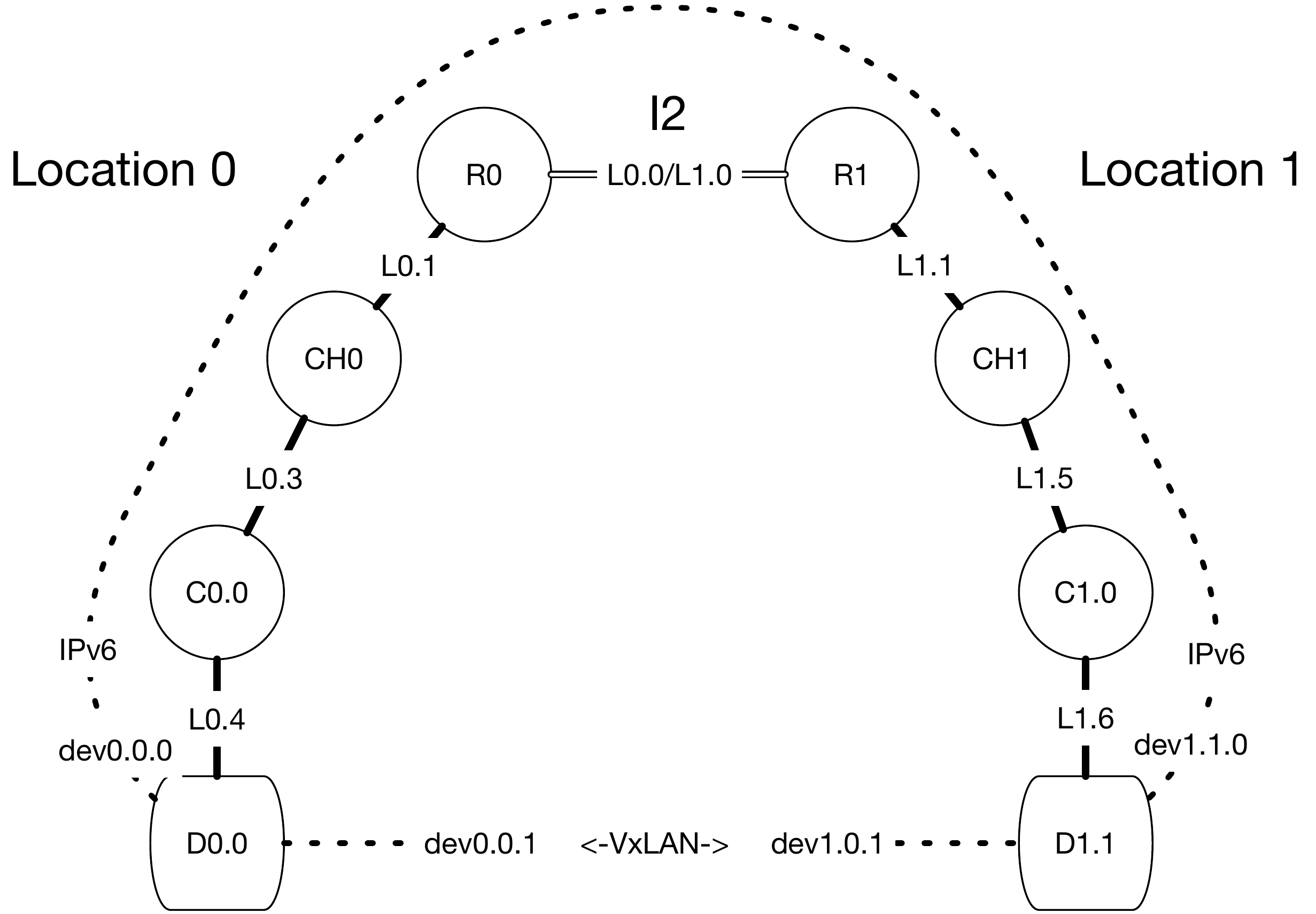}
  \caption{A tunnel between regions}
  \label{fig:cresco_topology_graph}
\end{figure}

Figure \ref{fig:sites_over_i2} shows two resource sites connected over a low-level connection through Internet2 (I2).  The link \emph{L0.0/L1.0} represents a Data-Link layer (L2) connection directly between edge routers \emph{R}.  Each edge site provides one or more computational resource providers\emph{CH} with optional storage \emph{S} resources.  Links \emph{LX.1} represent L2 communication between edge routers and computational/storage resources.  Container-provided or managed resources are represented by \emph{C}, where links \emph{LX.4 \& LX.6} represent several possible communication methods including, but not limited to, native IPv6 container endpoints, IPv4 tunnels over IPv6 networks between containers, or other protocols and transport mechanisms implemented in conjunction with \emph{CH} resources.  Devices, represented as \emph{D}, can be directly accessible globally or serve as data sources for edge gateways and/or higher-level processing functions.  Figure \ref{fig:cresco_topology_graph} provides an example of multi-transport communication between two endpoint devices managed by Cresco agents. 

\subsection{Large numbers of objects}  
\begin{figure}[h!]
  \centering
      \includegraphics[width=.22\textwidth]{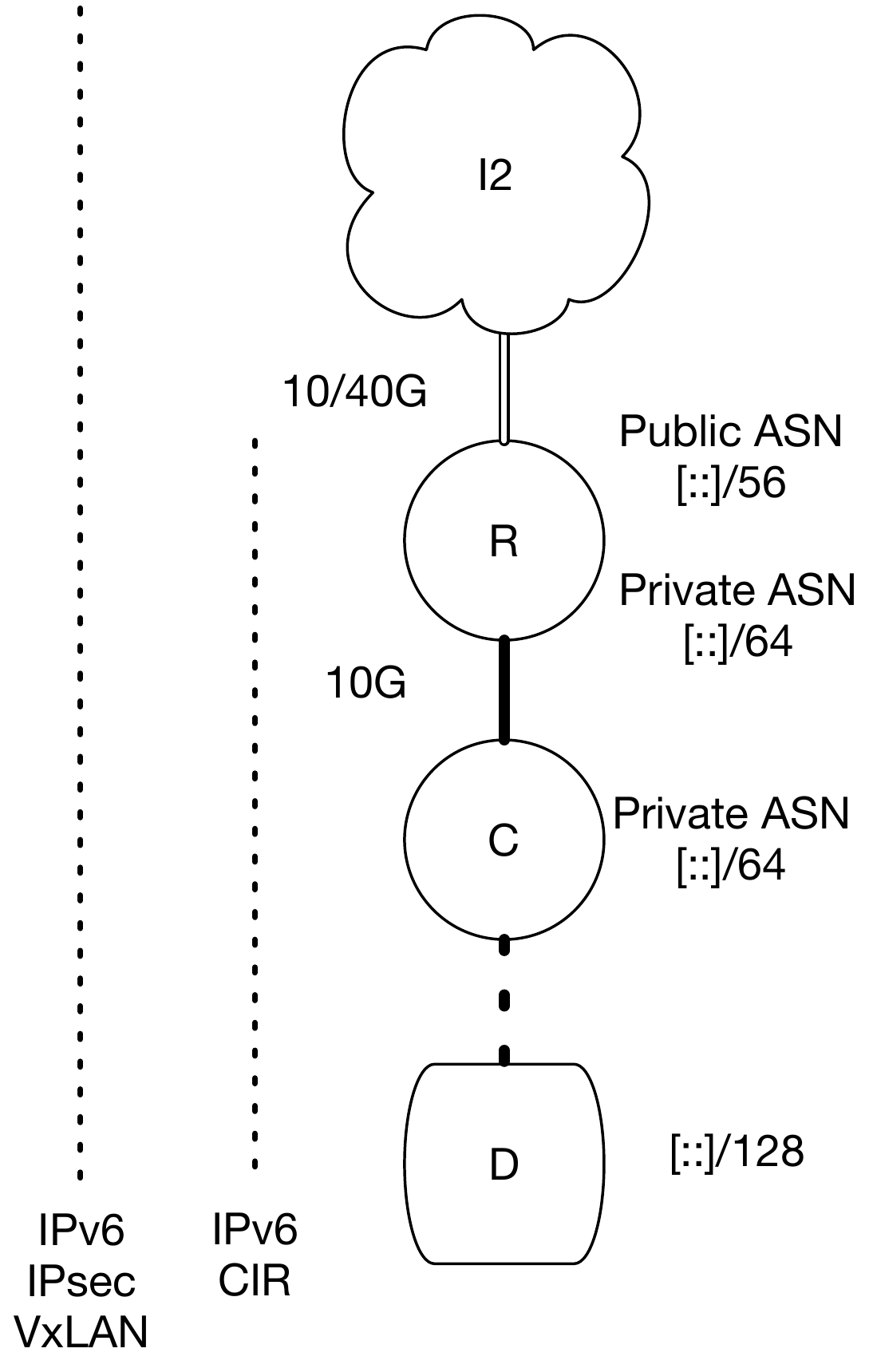}
  \caption{Local site (campus) network}
  \label{fig:site_address_topo}
\end{figure}

A critical step in managing large numbers of objects within a tested is to provide a low-level network to support object addressing and communications.  Considering there are over twice as many devices in China alone (9 billion as of 2014) as there are total IPv4 network addresses \cite{6851114}, we must carefully evaluate underlying management components.  GENI and many other testbeds are based on IPv4 addressing, which will not be sufficient to address a large number of devices.  We propose ``dual stack" operations allowing network communications on IPv4, IPv6, and other networks.  The control plane and associated network overlays should operate over IPv6 taking advantage of the QoS, security, and large address range features of the protocol.  Application layer services can operate over IPv4 or IPv6, depending on requirements.  IPv6 allows us to efficiently assign and route billions of addresses to individual edge and CPS resources.  In Figure \ref{fig:sites_over_i2}, routers designated as \emph{R0} and \emph{R1} are directly connected via L2 link.  These software\footnote{The software routers can be replaced by hardware routers if needed.} routers run Bird \cite{web:bird} Border Gateway Protocol (BGP) daemons and are capable of propagating IPv4/IPv6 routes between other sites and external networks.  Figure \ref{fig:site_address_topo} shows an example of a network that would extend from Internet2 (AL2S) to resources maintained with in a campus or city network.       

As shown in the previous figure, a \emph{/56} IPv6 range is advertised externally by the site.  For each compute host \emph{C} a \emph{/64} range is assigned providing $2^{64}$ addresses for each resource-providing or managing host.  Application or resource containers are identified under Cresco as functional units, which are maintained by Cresco agents.  Under test conditions, we have deployed hundreds of regional hosts under a global host, 10,000 Cresco Agents on a single regional host, and have provisioned as many as 10,000 function units on a single agent.  For each application container (functional unit) \emph{D} a \emph{/128} address is assigned, which allows the application container to natively communicate on IPv6 networks.  Additionally, \emph{/64} and \emph{/128} addresses can be assigned to external IoT gateways and directly to CPS devices.  A benefit of this architectural approach is that experimenters would have control of millions of distributed, yet managed objects associated with application functions, data end-points, physical devices, and other potentially addressable resources.          

\subsection{QoS Reservations and Enforcement} Testbeds are able to provide complex resource reservations but lack the global ability to enforce Service Level Commitments (SLC) for the resources provided.  In some cases, such as with I2 AL2S links \cite{web:al2sroadmap}, the underlying infrastructure does not support QoS controls necessary to satisfy SLCs.  However, through the use of Linux kernel namespace isolation and resource control groups (cgroup) \cite{rosen2013resource} we can manage computational and IO resources to a high degree of precision.  Table \ref{table:bandwidth_table} shows the degree of control provided by the kernel to limit network bandwidth from 10k - 100G across two large nodes connected over a 100G LAN.   

\begin{table}[h]
\begin{center}

\begin{tabular}{|l|l|l|l|l|}
\hline
{\bfseries Bandwidth Limit} & {\bfseries Send} & {\bfseries Receive} & {\bfseries Error}   & {\bfseries Unit}      \\ \hline
10              & 84.7 & 14.3    & 395.0\% & Kbits/sec \\ \hline
100             & 210  & 110     & 60.0\%  & Kbits/sec \\ \hline
1               & 1.13 & 1.01    & 7.0\%   & Mbits/sec \\ \hline
10              & 10.1 & 9.94    & 0.2\%   & Mbits/sec \\ \hline
100             & 99.6 & 99.3    & -0.6\%  & Mbits/sec \\ \hline
1000            & 981  & 981     & -1.9\%  & Mbits/sec \\ \hline
10              & 9.85 & 9.85    & -1.5\%  & Gbits/sec \\ \hline
100             & 38.8 & 38.8    & -61.2\% & Gbits/sec \\ \hline
\end{tabular}
\vspace{1mm}
\caption{Impact of Bandwidth Limits} 
\label{table:bandwidth_table}
\end{center}
\vspace{-7mm}
\end{table}

\begin{table}[h]
\begin{center}

\begin{tabular}{|l|l|l|l|l|}
\hline
{\bfseries +Latency ms} & {\bfseries Min}      & {\bfseries Avg}     & {\bfseries Max}     & {\bfseries Avg Error} \\ \hline
0           & 0.066    & 0.083   & 0.136    & 0.00\%    \\ \hline
1           & 1.076    & 1.117   & 1.144    & -3.40\%   \\ \hline
10          & 10.092   & 10.116  & 10.159   & -0.33\%   \\ \hline
100         & 100.094  & 100.123 & 100.193  & -0.04\%   \\ \hline
1000        & 1000.108 & 1000.14 & 1000.198 & -0.01\%   \\ \hline
\end{tabular}
\vspace{1mm}
\caption{Impact of Induced Latency} 
\label{table:latency_table}
\end{center}
\vspace{-8mm}
\end{table}

Likewise, Table \ref{table:latency_table} shows the introduction of latency using kernel control methods.  QoS policies can be implemented from routers to nodes and containers, as shown in Figure \ref{fig:site_address_topo}, as noted by \emph{IPv6 CIR} (Committed Information Rate).  As shown in the tables and figures, we have a high degree of network resources allowing for the simulation of various network topologies.  

Dynamic QoS operations can also be managed on the computational level to simulate the performance of various devices.  Table \ref{table:cpu_table} demonstrates the results of a simple single core benchmark \cite{pozo1999scimark} performed on several edge-focused devices including a Raspberry Pi, NVIDIA Nano, NVIDIA TX2, and other service-based systems.  The \emph{Native} column list the per millisecond average score over a five minute benchmark.  The \emph{Scale} indicates the ratio of a single \emph{Intel E7-4820 v4} core to allocate in order to simulate the other respective processors, based on observed native performance.  It is worth noting that an 0.1140 scaling factor was applied to all cases to account for overhead associated with the container-based testing.  The scaling factor was derived based on the differences between native and simulated results where the initial scaling factor for the \emph{Intel E7-4820 v4} processor was 1.  As shown in the table, we have a high degree of control on resources allowing for the simulation of various edge-focused devices.             

\begin{table}[h]
\begin{center}

\begin{tabular}{|l|l|l|l|l|}
\hline
{\bfseries CPU Type} & {\bfseries Native} & {\bfseries Scale}  & {\bfseries Simulated} & {\bfseries Error}  \\ \hline
Intel E7-4820 v4      & 44.17  & 1.1140 & 42.70     & -3.3\% \\ \hline
ARM Cortex-A57        & 19.73  & 0.4976 & 20.34     & 3.1\%  \\ \hline
AppliedMicro X-Gene 1 & 18.04  & 0.4548 & 19.63     & 8.8\%  \\ \hline
ARM Cortex-A57        & 14.58  & 0.3678 & 14.86     & 1.9\%  \\ \hline
ARM Cortex-A53        & 3.18   & 0.0802 & 2.88      & -9.6\% \\ \hline
Intel PHI (7200)      & 1.74   & 0.0439 & 1.59      & -8.8\% \\ \hline
\end{tabular}
\vspace{1mm}
\caption{Impact of Reduced CPU} 
\label{table:cpu_table}
\end{center}
\vspace{-7mm}
\end{table}

\subsection{End-to-end monitoring and measurement of resources}  End-to-end monitoring and measurement of federated resources used in experimentation and distributed applications is a challenge.  While high-level objects like provisioned network and compute resources are available, low-level monitoring and measurement of underlying edge resources and related networks are either not available or specific to underlying federations or resources.  For example, an application provisioned between two sites might use resources provided by different federated compute projects, with differing and possible unavailable low-level resource monitoring capabilities.  In addition, data related to the state of the physical network(s) providing connectivity between sites might also be unavailable.  The details pertaining to Cresco resource discovery, monitoring, measurement, and provisioning are beyond the scope of this paper.  However, Cresco agents have been used in resource-providing systems to verify operational status, including verification of SLCs.  In addition, lower-level infrastructure performance information is made available in conjunction with application-level performance information, allowing for the correlation of edge reservations to CPS application performance.  

\subsection{Application-level Specifications}

Application platform services, like those provided by public clouds, abstract the underlying details of infrastructure from application developers.  If an underlying infrastructure component fails on the platform, workload and data is reassigned to healthy resources.  For location independent applications, such as websites, a platform abstraction where the underlying service determines workload placement is attractive.  However, in the realm of edge-enabled CPS we want to selectively determine where workloads are assigned.  

In testbed and cloud computing infrastructure environments resource topologies are either requested as independent resource items or as collections of interconnected systems.  For instance, one might use Amazon EC2 to provision one or more independent virtual machines.  Likewise, a researcher might use GENI network to provision a multi-site topology connecting computational resources running specific software by means of the \emph{Rspec} \cite{faber2008resource} description language.  In both of these cases resources are described and provisioned statically through central control services.  Unlike cloud platforms, testbed provisioning systems typically do not detect and reassign resources on infrastructure failures.  In an edge-focused environment, infrastructure management must not only respond to application-level changes, it must anticipate, coordinate, and implement SLC-driven changes dynamically based on direct application interactions.  For instance, an overloaded edge at site \emph{A} must be able to intelligently interact with edge site \emph{B} and cloud site \emph{C} to determine appropriate workload offloading, based on observed workload characteristics. 

The proposed environment is based on a hierarchy of distributed agents.  Agents operate autonomously and are capable of dynamically developing operational topologies through an agent discovery processes.  Every agent can communicate with all other global agents through a protocol independent communication hierarchy.  Agent communication is restricted independently at each level of the hierarchy, based on group security policies.  Agents are responsible for application components and resources, Regional controllers are responsible for operations in their region, and global controllers manage regional controllers.  Request are filtered by input predicates and best-fit matching of resource to workload is pushed down to regional and agent-levels.    

Applications deployed on cloud platforms lack edge computing control and testbeds lack infrastructure abstractions and resiliency that simplifies the deployment of durable applications.  The aims should be to bring platform-like abstractions of infrastructure to edge-enabled CPS environments including testbeds.  While the details of the provisioning process are outside the scope of this document, as with the GENI Rspec, Cresco uses the CADL \cite{8480196} language to describe the desired application topology.  While Rspecs resource assignments are typically prescribed, but Cresco-managed resources are assigned through predicate filtering and best-fit matching.  For example, data collection services are pushed to specific locations as predicated by description, while higher-level processing can be assigned, and in the future reassigned, to an adjacent edge or cloud service.  In addition, current implementations make use of both public and private container registries.  Public registries are typically used to provide source containers for applications.  Private registries are used as both application sources and container snapshot targets.  The description of the application along with the ability to snapshot existing deployed applications allows Cresco to redeploy application components or entire topologies in the event of infrastructure failure.  In addition, where permitted by predicate assignment, workloads can be reassigned as environmental variables change.  For instance, location independent workloads on a specific edge can be migrated to cloud resources as additional local resources are needed.

\section{Related Work}

As previously mentioned, there are examples of IoT related edge computing on existing GENI and FIRE international testbeds.  While these IoT efforts might operate on international testbeds, the project focus it typically restricted to a smaller scale. \cite{Lima2019ExperimentalEF} compares 16 IoT or Wireless Sensor Network (WSN) testbeds and provides brief descriptions of their architecture.  The testbeds were selected for inclusion in the survey based on the number of citations the papers describing them had received. Of these 16, only one included more than 1000 nodes, though one (ExScal) claimed to support up to 10000 nodes. Many of the 16 testbeds do not support the higher-level functions like QoS for resources, end-to-end monitoring, M2M-focused operations, or durable high-level descriptions of applications.  \cite{Chernyshev2018InternetOT} includes a testbed not included in \cite{Lima2019ExperimentalEF}. \cite{Wu2018ASO} is a survey that covers several WSN testbeds. 

Table \ref{tab:relatedWork} summarizes information about a selection of testbeds from the works cited above. The ``Name" column lists the name of the testbed project. ``\# Devs." is the number of networked devices in the testbed and includes sensors, gateways, management nodes, and any other computers that participate in the network. The ``Scale" column refers to the physical, spatial scale of the testbed deployment as described in the literature. The ``App." describes what the testbed is used to test. The terms ``IoT" and ``WSN" may mean different things to different authors. We use ``IoT" to refer to systems with more high-level management and application-level capabilities and ``WSN" to refer to lower-level networks of sensors that connect wirelessly to a gateway using various forms of radio.

The ``Mgmt." column describes the system used to manage the testbed. The terms used are taken from \cite{Horneber2014}. The abbreviation used in table \ref{tab:relatedWork} is included next to each term in parentheses along with a short description.
\begin{itemize}
    \item Conventional network management (Conv.) - Techniques used for other computer networks like SNMP
    \item Management scripts (Scripts) - User-supplied scripts manage things like scheduling or access to nodes. Testbeds using this type of management often lack higher-level features.
    \item Management on sensor nodes (On nodes) - Sensors run distributed management software. These systems often have more high-level features
    \item Testbed interconnections (Interconn.) - Management system that supports interconnected testbeds. Supports higher-level features. 
\end{itemize}

\begin{table}[h]
    \centering
    \begin{tabular}{|l|c|l|l|l|}
    \hline
         \textbf{Name} & \textbf{\# Devs.} & \textbf{Scale} & \textbf{App.} & \textbf{Mgmt.} \\
         \hline
         Monarch Project & 9 & 700m x 300m & WSN & On nodes \\
         \hline
         MoteLab & 30 & Building & WSN & Scripts \\
         \hline
         Exscal & ~1200 & National & WSN & Unspecified \\
         \hline
         Kansei & 210 & Campus & IoT & Scripts \\
         \hline
         TWIST & 204 & Three floors & WSN & Scripts \\ 
         \hline
         Trio & 564 & 50 $\text{m}^2$ & IoT & On nodes\\
         \hline
         Mobile Emulab& 31 & 60 $\text{m}^2$ & IoT & Scripts \\
         \hline
         Citysense & 100  &  City & IoT & Scripts \\
         \hline
         Indirya & 127 & Building & IoT & Scripts \\
         \hline
         CONET & 26 & 528 $\text{m}^2$ & WSN & Scripts \\
         \hline
         Flocklab & 30 & Building & WSN & On nodes \\
         \hline
         Lazarescu & 1000 & ? & WSN & Conv.  \\
         \hline
         Smart Santander & 1373 & City & IoT & On nodes \\
         \hline
         IoT Lab & 1786 & Intnl. & IoT & Interconn. \\
         \hline
         JOSE & Unknown & National & IoT & Interconn. \\ 
         \hline
    \end{tabular}
    \vspace{1mm}
    \caption{IoT and WSN testbed comparison}
    \label{tab:relatedWork}
    \vspace{-7mm}
\end{table}

\section{Conclusions}

Existing global testbeds and cloud computing environments are not well suited to support edge-enabled CPS experimentation.  Most testbeds lack the production quality service aspects of public cloud computing offerings while commercial offerings often lack the edge computing resources and resource transparency offered by testbeds.  Both cloud and testbeds typically lack the ability to directly address, manage, and access very large numbers of devices.  In addition, neither testbeds or cloud offerings provide end-to-end monitoring, measurement, provisioning, and migration of services between edge and cloud resources.  Existing IoT efforts (Hubs, IoT-testbeds, etc) typically focus on low-level device communication and are limited to city or building-centric deployments.

We have presented potential methods to bridge the gap between existing global infrastructures, testbeds, IoT-centric commercial offerings efforts by addressing issues related to a) Stability of computing and network resources; b) Management of large number of objects; c) QoS enforcement of resource reservations; d) End-to-end monitoring and measurement of resources; e) Application-level experimentation specifications and deployment services.

\bibliographystyle{IEEEtran}
\bibliography{sources}

\end{document}